# Association of bone mineral density with reoperation rate following instrumented lumbar spinal fusion

Maximilian T. Löffler[1], MD, Niklas Loreck[1], Johannes Kaesmacher[1,2,3], MD, Felix Zibold[1,2], MD, Ehab Shiban[4], MD, Anna Rienmüller[4], MD, Martin Vazan[4], MD, Bernhard Meyer[4], MD, Yu-Mi Ryang[4]*, MD, Jan S. Kirschke[1]*, MD

1. Section of Neuroradiology, Department of Radiology, Klinikum rechts der Isar, Technische Universität München, Munich, Germany.
2. Institute of Diagnostic and Interventional Neuroradiology, University of Bern, Inselspital, Bern, Switzerland.
3. Department of Neurology, University of Bern, Inselspital, Bern, Switzerland.
4. Department of Neurosurgery, Klinikum rechts der Isar, Technische Universität München, Munich, Germany.

*Both senior authors contributed equally

**Corresponding Author**
Name:      Maximilian Löffler
Address:   Ismaninger Str. 22, 81675 Munich, Germany
Phone:     +49 89 4140 7639
Fax:       +49 89 4140 4887
Email:     m_loeffler@web.de

**Acknowledgements**
This project has received funding from the European Research Council (ERC) under the European Union's Horizon 2020 research and innovation program (grant agreement No 637164 — iBack — ERC-2014-STG).

# *Abstract*

Low bone mineral density (BMD) is believed to influence the outcome of instrumented spinal surgery and can lead to reoperation. Purpose of this observational and case-control study was to investigate the association of BMD with the risk of reoperation following instrumented lumbar spinal fusion (LSF).

For the observational study, 81 patients were included who received LSF with and without augmentation. For the case-control study, 18 patients who had reoperation following LSF were matched to 26 patients who did not have reoperation (matched by sex, age +/- 5 years, fused levels and PMMA-augmentation). Opportunistic BMD screening was performed in perioperative CT scans using asynchronous calibration. Mean BMD was compared between patients with and without reoperation in augmented and non-augmented surgeries.

In the observational study, prevalence of osteoporosis (BMD < 80 mg/cc) was 29% in non-augmented and 85% in augmented LSF. Seven of 48 patients with non-augmented (15%) and 4 of 33 patients with augmented LSF (12%) had reoperation. In non-augmented LSF, patients with reoperation had significantly lower BMD than patients without reoperation (P = 0.005). The best cut-off to predict reoperation after non-augmented LSF was BMD < 83.7 mg/cc. In the case-control study, patients with reoperation presented numerically lower BMD of 78.8 +/- 33.1 mg/cc than patients without reoperation with BMD of 89.4 +/- 39.7 mg/cc (P = 0.357).

Despite much lower BMD surgeries with PMMA-augmentation showed no higher reoperation rate compared to non-augmented surgeries. Patients with reoperation following LSF showed slightly lower BMD compared to matched patients without reoperation, but the difference was not statistically significant. Opportunistic BMD screening can be performed in preoperative CT, thus informing about osteoporotic bone, a potential risk factor of surgery failure.

**Abbreviations**

| | |
|---|---|
| BMD | bone mineral density |
| DXA | dual-energy X-ray absorptiometry |
| HU | Hounsfield unit |
| LSF | lumbar spinal fusion |
| MDCT | multidetector computed tomography |
| PMMA | polymethyl methacrylate |
| QCT | quantitative computed tomography |
| ROC | receiver operating characteristic |
| ROI | regions of interest |

# Introduction

Decreased bone mass is the main characteristic of osteoporosis leading to an increased risk of fractures [1]. Osteoporosis is the most common metabolic bone disease with 10% prevalence for people aged 50 years and older; the pre-stage (osteopenia) has a prevalence of 40% in the same age group [2]. The most common manifestation are osteoporotic vertebral compression fractures [3], which may require surgery with spinal fusion [4–6]. The indications for surgical stabilization are similar in osteoporotic and non-osteoporotic patients [7]. Osteoporosis is one predisposing factor for degenerative spine disease and microinstability. On the other hand, limited mobility due to spinal degeneration is a major predisposing factor for osteoporosis. Consequently, the prevalence of osteopenia or osteoporosis in patients undergoing spinal stabilization has been shown to be relatively high [7–9]. Vice versa, patients suffering from osteoporosis are more likely to receive surgical treatment by spinal fusion.

Studies over the last three decades continually report failure rates of 13–19% for instrumented fusion of the lumbar spine [10–13]. There is limited evidence from *in vivo* studies that associate low bone density with an increased risk of complications and surgical failure rates [14,15], whereas many biomechanical studies concerning this topic exist [16–20]. Surgical failure may be due to impaired screw fixation or purchase [14,21,22], interbody cage subsidence [23], junctional kyphosis adjacent to the instrumented levels [24,25] or reduced osteogenic potential [26]. In summary, osteoporosis is believed to be an independent risk factor for instrumentation failure [27], and hence for short- and long-term revision surgery. Therefore, successful instrumented surgery in the osteoporotic spine is especially challenging. Preoperative bone mineral density (BMD) assessment objectifies doubts about bone substance, thus allowing to acknowledge this challenge in the surgical planning process. A survey among spine surgeons showed that prior to instrumented fusion only 44% routinely obtained dual-energy X-ray absorptiometry (DXA) examinations if osteoporosis was suspected [27]. Supplementary DXA examinations may become obsolete, if volumetric BMD can be opportunistically evaluated on preoperative CT imaging. The feasibility and validity of opportunistic quantitative CT (QCT) has been extensively shown [22,28–34]. For reasons of convenience BMD will refer to volumetric density throughout this text, if not stated otherwise.

We carried out a retrospective observational study and case-control study to investigate whether reduced BMD is associated with an increased rate of reoperations following elective lumbar spinal fusion (LSF) and if this association is dependent on the use of polymethyl methacrylate (PMMA)-augmentation.

# Methods

**Retrospective study**

This retrospective analysis of patient data was approved by the local institutional review board. We reviewed 1441 consecutive patients, who underwent a neurosurgical operation involving the lumbar spine in our institution in the years 2010 to 2014. We only included patients, who had primary elective LSF with either a non-augmented rigid pedicle screw-rod system (Pangea Degenerative Spine System; Synthes, West Chester, PA, USA) or a rigid system with PMMA-augmented pedicle screw fixation. We excluded patients with vertebral neoplasia, without CT and with non-elective surgeries. Indications for elective LSF were related to degenerative spine disease (spondylolisthesis, spondylolysis or spinal stenosis). Following this algorithm 33 patients with PMMA-augmented surgeries and 48 patients with non-augmented surgeries were identified (Figure 1). Sociodemographic data and information about index and revision surgeries were extracted from patient files and operation reports (Table 1). Seven patients with non-augmented LSF and 4 patients with augmented LSF encountered complications which led to revision surgery (Table 2, Figure 2). Reoperations after immediate surgery related complications, such as misplaced pedicle screws in 2 cases, were not taken into account.

**Case-control study**

Eighteen patients with reoperation following LSF were matched by sex, age ± 5 years, and fused levels to 26 patients without reoperation. Again, we only included patients, who had primary elective LSF for indication related to degenerative spine disease (spondylolisthesis, spondylolysis or spinal stenosis) with either a non-augmented rigid pedicle screw-rod system or a rigid system with PMMA pedicle screw augmentation. In the hospital's records, patients without reoperation did not present sensory-motor deficit or severe pain at last visit after a median follow-up of 154 days (range 5 – 2183).

**MDCT image acquisition**

Pre-operative or immediate post-operative CT scans were used for opportunistic BMD screening. CT scans were performed on three multidetector computed tomography (MDCT) scanners in the same hospital (Philips Brilliance 64, Philips Medical Care; Siemens Somatom Definition AS+ and Definition AS, Siemens Healthineers), partly with administration of intravenous contrast medium (Imeron 400, Bracco). Image data was acquired in helical mode with a peak tube voltage of 120 kVp for standard and 140kVp for postmyelography studies.

**Opportunistic BMD screening**

Volumetric BMD (in mg/cm³) in trabecular bone of at least one lumbar vertebra was opportunistically assessed. Therefore, X-ray attenuation in Hounsfield units (HU) had to be converted to BMD. HU-to-BMD conversion equations were calculated by asynchronous calibration as reported in a previous study [35]. For postmyelography studies with 140kVp tube voltage, another previously reported conversion equation was used [22]. For contrast-enhanced CT scans, BMD correction offsets for arterial (–8.6 mg/cm³) and portal-venous contrast phase (–15.8 mg/cm³) were added based on previous investigations [36]. HU was measured with tools of the institutional picture archiving and communication system (PACS) software (Sectra IDS7; Sectra AB, Linköping, Sweden). At first, additional sagittal stacks of 15 mm thickness (increment 2 mm) were calculated to average attenuation signals. Herein average HUs were extracted from circular regions of interest (ROIs) in the midsagittal plane, placed by an experienced radiologist in the cancellous bone of at least one lumbar vertebral body. Fractured vertebra or those with apparent alterations of the cancellous bone due to degeneration or hemangioma were omitted. ROIs spanning approximately half of the vertebral height in diameter were vertically centered with equal distance to cortical bone and ventrally aligned [22]. Following the ACR practice parameters for bone densitometry, osteoporosis was defined as $BMD < 80$ mg/cm$^3$ and osteopenia as $80$ mg/cm$^3 \leq BMD \leq 120$ mg/cm$^3$ [37].

**Statistical analysis**

Means were compared with independent sample t-tests assuming equality of variances depending on Levene's test. In a receiver operating characteristic (ROC) analysis, the classification performance of BMD to predict reoperations was tested. BMD thresholds were determined with maximum Youden index. Statistical analyses were conducted with IBM SPSS Statistics 24 (IBM Corp., Armonk, NY, USA). Level of significance for all tests was defined as $P < 0.05$.

## *Results*

**Prevalence of low BMD in retrospective study**

Mean BMD was significantly lower in PMMA-augmented surgeries with 60.2 ± 24.1 mg/cm³ than in non-augmented surgeries with 104.8 ± 37.9 mg/cm³ (P = 0.002; Table 1). The length of the fixation constructs differed significantly with a median of 2 fused segments (mean 2.3) in PMMA-augmented versus a median of 1 fused segment (mean 1.4) in non-augmented surgeries (P < 0.001; Table 1). There was no significant difference in mean BMD between men with 94.4 ± 34.5 mg/cm³ and women with 81.3 ± 42.2 mg/cm³ (P = 0.142). Patients with non-augmented LSF who underwent reoperation had significantly lower mean BMD of 73.0 ± 18.4 mg/cm³ than those who did not with a mean BMD of 110.2 ± 37.8 mg/cm³ (P = 0.015; Table 3). There was no significant difference in BMD between patients with PMMA-augmented LSF who underwent reoperation and those who did not (P = 0.621). The best threshold to predict reoperation in non-augmented LSF according to ROC analysis was at a BMD below 83.7 mg/cm³ (area under the ROC curve A = 0.798; 95% confidence interval [CI] = 0.649–0.946; P = 0.013; Table 3). As the difference of BMD between augmented surgeries with and without reoperation was not significant, ROC analysis and logistic regression was not performed for this subgroup.

**Reoperations in patients with non-osteoporotic BMD in case-control study**

In the case-control study, 18 patients with reoperation presented similarly low BMD of 78.8 ± 33.1 mg/cm³ compared to 26 matched patients without reoperation with BMD of 89.4 ± 39.7 mg/cm³ (Table 4). This numerical difference was not statistically significant (P = 0.357).

## *Discussion*

The observational study showed that opportunistic BMD assessment on preoperative CT allows detection of osteoporotic bone density in patients who are scheduled for LSF. The prevalence of osteoporosis in this elderly group of patients undergoing elective surgery is relatively high. We were able to demonstrate that patients with PMMA-augmented LSF exhibited much lower BMD than those with non-augmented LSF, while both groups showed an almost equal rate of reoperations. The case-control study showed that lumbar BMD was similar in patients with reoperation compared to matched controls, who did not have reoperation. Despite not being statistically significant, BMD was numerically lower in cases with reoperations than in controls without reoperation. However, several limitations of these studies have to be addressed.

Results indicated here are in line with previous studies. Low BMD assessed using HU measurements on preoperative CT scans has been associated with adjacent vertebral body fractures after spinal fusion surgery [15]. In a case-control study decreased HU on preoperative scans were associated with symptomatic pseudarthrosis on one-year follow-up after spine surgery with posterolateral lumbar fusion [38]. Patients with radiographic signs of screw loosening and non-fusion on follow-up after LSF, had lower BMD than those without these signs [14]. Another study showed that BMD measurements by opportunistic QCT adequately differentiated patients with and without osteoporotic fractures and could predict incidental fractures and screw loosening after spinal fusion [22]. Decreased BMD in the cervical spine was identified as the major predisposing factor for the occurrence of traumatic odontoid fractures in elderly patients [32].

Given that patients in our observational study with PMMA-augmented LSF had significantly lower BMD than those without augmentation, it is noteworthy that these patients did not show a higher reoperation rate. PMMA-augmented screw fixation is recommended in osteoporotic bone [4,5], because it improves the fixation and fatigue strength *ex vivo* [39], reduces the risk of screw loosening and pullout [40], and increases fusion rates with maintained correction angles *in vivo* [41]. Accordingly, reviews advocate PMMA-augmented screw fixations and other technical modifications like long-segment constructs to reduce the risk of instrumentation failure in osteoporosis patients [4,42,43]. Multiple points of fixation have been recommended in the osteoporotic spine for a long time [44]. To avoid ending within a spinal transition zone or a kyphotic section long-segment constructs seem beneficial [25,45,46], as these regions are typically prone to adjacent segment degeneration, adjacent vertebral body fractures or implant failure. As patients in our study with augmented surgeries had also one more fused segments on average, there might be a positive effect of both these factors leading to an even slightly decreased reoperation rate (12% vs. 15%), despite substantially lower BMD.

In the observational study, patients with elective LSF had a mean age of 67 years and showed a 52% prevalence of osteoporosis. Women had slightly, but non-significantly lower BMD than men. There was a predominance of women in the group with PMMA-augmented surgeries (79%) showing substantially lower BMD compared to non-augmented surgeries, which is probably due to postmenopausal changes in bone metabolism. When only looking at the patients who received non-augmented surgeries, the prevalence of osteoporotic BMD was still 29%. The high prevalence of osteoporosis may be due the high mean age of our study population. However indications for surgery in elderly patients persist, despite an increased overall surgical risk, because favorable outcomes after successful spinal fusion can be obtained in the majority of these patients [47]. Previously, a nearly 30% prevalence of osteoporotic BMD or fragile bone strength has been reported in women between the age of 50 to 70 years undergoing spinal fusion [8]. However, equally high rates of osteoporotic BMD were observed in the over 50 year old population for men and women prior to spinal fusion surgery [7,9]. Thus, biomechanical considerations and use of the aforementioned surgical techniques are of increasing importance when performing spinal instrumentations in the osteoporotic spine [43,46].

Of note, BMD measurements are not performed on a regular basis prior to surgery in our hospital. We employed the method of assessing lumbar BMD on routine CT imaging, termed opportunistic QCT [31], which has been validated [28] and applied in various studies [22,29,30,32–34,48], showing low short- and long-term reproducibility errors [30]. Previous studies demonstrated that lumbar BMD can be assessed in sagittal reformations of contrast-enhanced MDCT and used to differentiate patients with and without osteoporotic fractures [30], as well as predict these fractures [29]. Linear correction equations can be adjusted for systematic bias of apparent bone density related to different calibration techniques and contrast application [49]. We analyzed CT images obtained on different devices for indications other than densitometry and applied asynchronous calibration to calculate BMD [22,35]. In contrast to direct HU measurements, which are CT device dependent, calibration ensures inter-scanner and/or inter-study comparability of absolute BMD values. Moreover, predefined BMD thresholds for osteoporosis can be used [37]. In order to determine whether there is an increased risk of complications and reoperation after LSF, we estimated a BMD threshold of 83.7 mg/cm³ for patients with non-augmented surgeries in our study. In a biomechanical study, BMD of less than 80 mg/cm³ was associated with early screw loosening and unsatisfactory spinal fixation [19]. Okuyama et al. hypothesized that an areal BMD below 0.674 ± 0.104 g/cm² indicated a potentially increased risk of spinal fusion failure [14]. Although difficult to compare to volumetric BMD, this value certainly lies within the osteoporotic range. Apparently, the estimated cutoff in our study matched closely with the proposed threshold of lumbar BMD for the diagnosis of osteoporosis [37]. This emphasizes the importance of assessing lumbar BMD prior to spinal instrumentation.

**Limitations**

The presented studies are based on retrospective data and therefore prone to bias. Loss of follow-up is a major confounding factor in this observational study, though expected to be similar across groups. Moreover, there are no objective criteria for reoperation. The decision to have revision surgery is inherently subjective and influenced by the patient's and surgeon's preferences. The authors are aware that sagittal balance is an important biomechanical factor of the spine and has an impact on the outcome of LSF. Unfortunately, radiographs that allowed analyzing sagittal vertebral axis before and after surgery were not available for the presented data, since they were not part of routine perioperative workup in our institution until 2015.

## *Conclusion*

Despite much lower BMD surgeries with PMMA-augmentation showed no higher reoperation rate compared to non-augmented surgeries, which could be explained by the improved pedicle screw purchase through augmentation. Patients with reoperation following LSF showed slightly lower BMD compared to matched patients without reoperation, but the difference was not statistically significant. Potential loss of follow-up and the lack of objectivity in the decision to undergo reoperation have to be recognized as major limitations of the presented results. However, opportunistic BMD evaluation is feasible and can be advised before LSF, thus informing about osteoporotic bone.

## *References*

[1]  VA. Consensus development conference: Diagnosis, prophylaxis, and treatment of osteoporosis. Am J Med 1993;94:646–50. https://doi.org/10.1016/0002-9343(93)90218-E.

[2]  Wright NC, Looker AC, Saag KG, Curtis JR, Delzell ES, Randall S, et al. The recent prevalence of osteoporosis and low bone mass in the United States based on bone mineral density at the femoral neck or lumbar spine. J Bone Miner Res 2014;29:2520–6. https://doi.org/10.1002/jbmr.2269.

[3]  Kammerlander C, Zegg M, Schmid R, Gosch M, Luger TJ, Blauth M. Fragility Fractures Requiring Special Consideration. Clinics in Geriatric Medicine 2014;30:361–72. https://doi.org/10.1016/j.cger.2014.01.011.

[4]  Heini PF. The current treatment--a survey of osteoporotic fracture treatment. Osteoporotic spine fractures: the spine surgeon's perspective. Osteoporos Int 2005;16 Suppl 2:S85-92. https://doi.org/10.1007/s00198-004-1723-1.

[5]  Krappinger D, Kastenberger TJ, Schmid R. [Augmented posterior instrumentation for the treatment of osteoporotic vertebral body fractures]. Oper Orthop Traumatol 2012;24:4–12. https://doi.org/10.1007/s00064-011-0098-7.

[6]  Patil S, Rawall S, Singh D, Mohan K, Nagad P, Shial B, et al. Surgical patterns in osteoporotic vertebral compression fractures. Eur Spine J 2013;22:883–91. https://doi.org/10.1007/s00586-012-2508-4.

[7]  Chin DK, Park JY, Yoon YS, Kuh SU, Jin BH, Kim KS, et al. Prevalence of osteoporosis in patients requiring spine surgery: Incidence and significance of osteoporosis in spine disease. Osteoporosis International 2007;18:1219–24. https://doi.org/10.1007/s00198-007-0370-8.

[8]  Burch S, Feldstein M, Hoffmann PF, Keaveny TM. Prevalence of Poor Bone Quality in Women Undergoing Spinal Fusion Using Biomechanical-CT Analysis. SPINE 2016;41:246–52. https://doi.org/10.1097/BRS.0000000000001175.

[9]  Wagner SC, Formby PM, Helgeson MD, Kang DG. Diagnosing the Undiagnosed: Osteoporosis in Patients Undergoing Lumbar Fusion. Spine 2016;41:E1279–83. https://doi.org/10.1097/BRS.0000000000001612.

[10] Greiner-Perth R, Boehm H, Allam Y, Elsaghir H, Franke J. Reoperation rate after instrumented posterior lumbar interbody fusion: a report on 1680 cases. Spine 2004;29:2516–20.

[11] Irmola TM, Häkkinen A, Järvenpää S, Marttinen I, Vihtonen K, Neva M. Reoperation Rates Following Instrumented Lumbar Spine Fusion. Spine 2017. https://doi.org/10.1097/BRS.0000000000002291.

[12] Malter AD, McNeney B, Loeser JD, Deyo RA. 5-year reoperation rates after different types of lumbar spine surgery. Spine 1998;23:814–20.

[13] Martin BI, Mirza SK, Comstock BA, Gray DT, Kreuter W, Deyo RA. Reoperation rates following lumbar spine surgery and the influence of spinal fusion procedures. Spine 2007;32:382–7. https://doi.org/10.1097/01.brs.0000254104.55716.46.

[14] Okuyama K, Abe E, Suzuki T, Tamura Y, Chiba M, Sato K. Influence of bone mineral density on pedicle screw fixation: a study of pedicle screw fixation augmenting posterior lumbar interbody fusion in elderly patients. Spine J 2001;1:402–7.

[15] Meredith DS, Schreiber JJ, Taher F, Cammisa FP, Girardi FP. Lower preoperative Hounsfield unit measurements are associated with adjacent segment fracture after spinal fusion. Spine 2013;38:415–8. https://doi.org/10.1097/BRS.0b013e31826ff084.

[16] Konstantinidis L, Helwig P, Hirschmüller A, Langenmair E, Südkamp NP. When is the stability of a fracture fixation limited by osteoporotic bone? Injury 2016;47:S27–32. https://doi.org/10.1016/S0020-1383(16)47005-1.

[17] Paxinos O, Tsitsopoulos PP, Zindrick MR, Voronov LI, Lorenz MA, Havey RM, et al. Evaluation of pullout strength and failure mechanism of posterior instrumentation in normal and osteopenic thoracic vertebrae. J Neurosurg Spine 2010;13:469–76. https://doi.org/10.3171/2010.4.SPINE09764.

[18] Knöller SM, Meyer G, Eckhardt C, Lill CA, Schneider E, Linke B. Range of motion in reconstruction situations following corpectomy in the lumbar spine: a question of bone mineral density? Spine 2005;30:E229-235.

[19] Wittenberg RH, Shea M, Swartz DE, Lee KS, White AA, Hayes WC. Importance of bone mineral density in instrumented spine fusions. Spine 1991;16:647–52.

[20] Eysel P, Schwitalle M, Oberstein A, Rompe JD, Hopf C, Küllmer K. Preoperative estimation of screw fixation strength in vertebral bodies. Spine 1998;23:174–80.

[21] Reitman CA, Nguyen L, Fogel GR. Biomechanical evaluation of relationship of screw pullout strength, insertional torque, and bone mineral density in the cervical spine. J Spinal Disord Tech 2004;17:306–11.

[22] Schwaiger BJ, Gersing AS, Baum T, Noel PB, Zimmer C, Bauer JS. Bone Mineral Density Values Derived from Routine Lumbar Spine Multidetector Row CT Predict Osteoporotic Vertebral Fractures and Screw Loosening. American Journal of Neuroradiology 2014;35:1628–33. https://doi.org/10.3174/ajnr.A3893.

[23] Oh KW, Lee JH, Lee J-H, Lee D-Y, Shim HJ. The Correlation Between Cage Subsidence, Bone Mineral Density, and Clinical Results in Posterior Lumbar Interbody Fusion. Clin Spine Surg 2017;30:E683–9. https://doi.org/10.1097/BSD.0000000000000315.

[24] Wang H, Ma L, Yang D, Wang T, Yang S, Wang Y, et al. Incidence and risk factors for the progression of proximal junctional kyphosis in degenerative lumbar scoliosis following long instrumented posterior spinal fusion. Medicine 2016;95:e4443. https://doi.org/10.1097/MD.0000000000004443.

[25] DeWald CJ, Stanley T. Instrumentation-related complications of multilevel fusions for adult spinal deformity patients over age 65: surgical considerations and treatment options in patients with poor bone quality. Spine 2006;31:S144-51. https://doi.org/10.1097/01.brs.0000236893.65878.39.

[26] Kim B-H, Jung H-G, Park K-H, Kim D-H, Choi Y-S. The Effectiveness of Bone Mineral Density as Supplementary Tool for Evaluation of the Osteogenic Potential in Patients with Spinal Fusion. Asian Spine Journal 2009;3:1. https://doi.org/10.4184/asj.2009.3.1.1.

[27] Dipaola CP, Bible JE, Biswas D, Dipaola M, Grauer JN, Rechtine GR. Survey of spine surgeons on attitudes regarding osteoporosis and osteomalacia screening and treatment for fractures, fusion surgery, and pseudoarthrosis. Spine J 2009;9:537–44. https://doi.org/10.1016/j.spinee.2009.02.005.

[28] Bauer JS, Henning TD, Müeller D, Lu Y, Majumdar S, Link TM. Volumetric Quantitative CT of the Spine and Hip Derived from Contrast-Enhanced MDCT: Conversion Factors. American Journal of Roentgenology 2007;188:1294–301. https://doi.org/10.2214/AJR.06.1006.

[29] Baum T, Müller D, Dobritz M, Wolf P, Rummeny EJ, Link TM, et al. Converted lumbar BMD values derived from sagittal reformations of contrast-enhanced MDCT predict incidental osteoporotic vertebral fractures. Calcif Tissue Int 2012;90:481–7. https://doi.org/10.1007/s00223-012-9596-3.

[30] Baum T, Müller D, Dobritz M, Rummeny EJ, Link TM, Bauer JS. BMD measurements of the spine derived from sagittal reformations of contrast-enhanced MDCT without dedicated software. Eur J Radiol 2011;80:e140-145. https://doi.org/10.1016/j.ejrad.2010.08.034.

[31] Engelke K. Quantitative Computed Tomography-Current Status and New Developments. J Clin Densitom 2017;20:309–21. https://doi.org/10.1016/j.jocd.2017.06.017.

[32] Kaesmacher J, Schweizer C, Valentinitsch A, Baum T, Rienmüller A, Meyer B, et al. Osteoporosis Is the Most Important Risk Factor for Odontoid Fractures in the Elderly. J Bone Miner Res 2017;32:1582–8. https://doi.org/10.1002/jbmr.3120.

[33] Link TM, Koppers BB, Licht T, Bauer J, Lu Y, Rummeny EJ. In vitro and in vivo spiral CT to determine bone mineral density: initial experience in patients at risk for osteoporosis. Radiology 2004;231:805–11. https://doi.org/10.1148/radiol.2313030325.

[34] Papadakis AE, Karantanas AH, Papadokostakis G, Petinellis E, Damilakis J. Can abdominal multi-detector CT diagnose spinal osteoporosis? Eur Radiol 2009;19:172–6. https://doi.org/10.1007/s00330-008-1099-2.

[35] Löffler MT, Jacob A, Valentinitsch A, Rienmüller A, Zimmer C, Ryang Y-M, et al. Improved prediction of incident vertebral fractures using opportunistic QCT compared to DXA. Eur Radiol 2019. https://doi.org/10.1007/s00330-019-06018-w.

[36] Kaesmacher J, Liebl H, Baum T, Kirschke JS. Bone Mineral Density Estimations From Routine Multidetector Computed Tomography: A Comparative Study of Contrast and Calibration Effects. J Comput Assist Tomogr 2017;41:217–23. https://doi.org/10.1097/RCT.0000000000000518.

[37] American College of Radiology. ACR-SPR-SSR practice parameter for the performance of quantitative computed tomography (QCT) bone densitometry 2014.

[38] Nguyen HS, Shabani S, Patel M, Maiman D. Posterolateral lumbar fusion: Relationship between computed tomography Hounsfield units and symptomatic pseudoarthrosis. Surg Neurol Int 2015;6:S611-614. https://doi.org/10.4103/2152-7806.170443.

[39] Burval DJ, McLain RF, Milks R, Inceoglu S. Primary pedicle screw augmentation in osteoporotic lumbar vertebrae: biomechanical analysis of pedicle fixation strength. Spine 2007;32:1077–83. https://doi.org/10.1097/01.brs.0000261566.38422.40.

[40] Frankel BM, Jones T, Wang C. Segmental polymethylmethacrylate-augmented pedicle screw fixation in patients with bone softening caused by osteoporosis and metastatic tumor involvement: a clinical evaluation. Neurosurgery 2007;61:531–7; discussion 537-538. https://doi.org/10.1227/01.NEU.0000290899.15567.68.

[41] Sawakami K, Yamazaki A, Ishikawa S, Ito T, Watanabe K, Endo N. Polymethylmethacrylate augmentation of pedicle screws increases the initial fixation in osteoporotic spine patients. J Spinal Disord Tech 2012;25:E28-35. https://doi.org/10.1097/BSD.0b013e318228bbed.

[42] Fischer CR, Hanson G, Eller M, Lehman RA. A Systematic Review of Treatment Strategies for Degenerative Lumbar Spine Fusion Surgery in Patients With Osteoporosis. Geriatr Orthop Surg Rehabil 2016;7:188–96. https://doi.org/10.1177/2151458516669204.

[43] Lehman RA, Kang DG, Wagner SC. Management of osteoporosis in spine surgery. J Am Acad Orthop Surg 2015;23:253–63. https://doi.org/10.5435/JAAOS-D-14-00042.

[44] Hu SS. Internal fixation in the osteoporotic spine. Spine 1997;22:43S-48S.

[45] Dodwad S-NM, Khan SN. Surgical stabilization of the spine in the osteoporotic patient. Orthop Clin North Am 2013;44:243–9. https://doi.org/10.1016/j.ocl.2013.01.008.

[46] Ponnusamy KE, Iyer S, Gupta G, Khanna AJ. Instrumentation of the osteoporotic spine: biomechanical and clinical considerations. Spine J 2011;11:54–63. https://doi.org/10.1016/j.spinee.2010.09.024.

[47] Okuda S, Oda T, Miyauchi A, Haku T, Yamamoto T, Iwasaki M. Surgical outcomes of posterior lumbar interbody fusion in elderly patients. J Bone Joint Surg Am 2006;88:2714–20. https://doi.org/10.2106/JBJS.F.00186.

[48] Engelke K, Lang T, Khosla S, Qin L, Zysset P, Leslie WD, et al. Clinical Use of Quantitative Computed Tomography-Based Advanced Techniques in the Management of Osteoporosis in Adults: the 2015 ISCD Official Positions-Part III. J Clin Densitom 2015;18:393–407. https://doi.org/10.1016/j.jocd.2015.06.010.

[49] Kaesmacher J, Liebl H, Baum T, Kirschke JS. Bone Mineral Density Estimations From Routine Multidetector Computed Tomography: A Comparative Study of Contrast and Calibration Effects. J Comput Assist Tomogr 2016. https://doi.org/10.1097/RCT.0000000000000518.

# Tables

Table 1: Patients' characteristics in the observational study stratified according to the surgical technique. *Posterolateral interbody fusion with synthetic bone graft was performed. ALIF, anterior lumbar interbody fusion; PLIF, posterior lumbar interbody fusion; TLIF, transforaminal lumbar interbody fusion; XLIF, extreme lateral interbody fusion.

|  |  | PMMA-augmented n = 33 | Non-augmented n = 48 | PMMA- vs. non-augmented | All n = 82 |
|---|---|---|---|---|---|
| Women, n (%) |  | 26 (79%) | 22 (46%) | P = 0.002 | 48 (59%) |
| Age, yrs, mean (range) |  | 76 (51–89) | 61 (30–84) | P < 0.001 | 67 (30–89) |
| BMD, mg/cm³, mean (SD) |  | 60.2 (24.1) | 104.8 (37.9) | P < 0.001 | 86.6 (39.5) |
| Osteoporosis, n (%) |  | 28 (85%) | 14 (29%) | P < 0.001 | 42 (52%) |
| Osteopenia, n (%) |  | 5 (15%) | 19 (40%) | P = 0.012 | 24 (30%) |
| Duration of surgery, min (range) |  | 228 (95–403) | 213 (77–385) | P = 0.359 | 219 (77–403) |
| Fused segments, n (%) | 1 | 9 (27%) | 33 (69%) | P < 0.001 | 42 (52%) |
|  | 2 | 11 (33%) | 11 (23%) |  | 22 (27%) |
|  | 3 | 7 (21%) | 3 (6%) |  | 10 (12%) |
|  | 4 | 5 (15%) | 1 (2%) |  | 6 (7%) |
|  | 5 | 1 (3%) | 0 |  | 1 (1%) |
| Segment L5/S1, n (%) | Included | 19 (58%) | 33 (69%) | P = 0.309 | 52 (64%) |
|  | Above | 14 (42%) | 15 (31%) |  | 29 (36%) |
| Interbody fusion type, n (%) | TLIF/PLIF | 21 (64%) | 38 (79%) | P = 0.086 | 59 (73%) |
|  | XLIF | 3 (9%) | 2 (4%) |  | 5 (6%) |
|  | ALIF | 3 (9%) | 7 (15%) |  | 10 (12%) |
|  | No cage* | 6 (18%) | 1 (2%) |  | 7 (9%) |
| Reoperations, n (%) |  | 4 (12%) | 7 (15%) | P = 0.754 | 11 (14%) |

Table 2: Complications after lumbar spinal fusion in the observational study stratified according to the surgical technique. *Reoperations after these complications were not taken into account.

|  | PMMA-augmented | | Non-augmented | |
|---|---|---|---|---|
| Complication | N | Reoperation interval, mean (range) | N | Reoperation interval, mean (range) |
| Instrumentation failure | 1 | 13.1 months | 3 | 21.0 (9.4–43.7) months |
| Adjacent segment degeneration | 2 | 22.7 (17.7–27.6) months | 3 | 39.9 (18.7–52.8) months |
| New fracture | 1 | 2.4 months | 1 | 22.4 months |
| Misplaced pedicle screw |  |  | 2* | 13 (2–24) days |

Table 3: Mean BMD of patients with/without reoperation in the observational study stratified according to the surgical technique. *ROC analysis was not performed in absence of significant BMD difference between PMMA-augmented surgeries with/without reoperation. AUC, area under the ROC curve.

|  | Group size | BMD, mean, mg/cm³ (SD) | No reoperation vs. reoperation | ROC AUC (CI, Sig.) | BMD threshold, mg/cm³ (Youden index) |
|---|---|---|---|---|---|
| Non-augmented without reoperation | 41 | 110.2 (37.8) | P = 0.015 | 0.798 (0.649–0.946, P = 0.013) | 83.7 (J = 0.66) |
| Non-augmented with reoperation | 7 | 73.0 (18.4) | | | |
| PMMA-augmented without reoperation | 29 | 61.0 (24.9) | P = 0.621 | * | |
| PMMA-augmented with reoperation | 4 | 54.5 (19.3) | | | |

Table 4: Patients' characteristics in the case-control study.

|  |  | Case n = 18 | Control n = 26 | Case vs. control |
|---|---|---|---|---|
| Age, yrs, mean (SD) | | 68.5 (9.5) | 69.5 (8) | P = 0.71 |
| BMD, mg/cm³, mean (SD) | | 78.8 (33.1) | 89.4 (39.7) | P = 0.359 |
| PMMA-augmented, n | | 3 | 5 | P = 0.828 |
| Fused segments, n | 1 | 6 | 8 | P = 0.858 |
| | 2 | 8 | 12 | P = 0.911 |
| | 3 | 3 | 5 | P = 0.828 |
| | 4 | 1 | 1 | P = 0.789 |
| Non-enhanced CT, n | | 15 | 20 | P = 0.604 |
| Postmyelography CT, n | | 1 | 3 | P = 0.497 |

## *Figures*

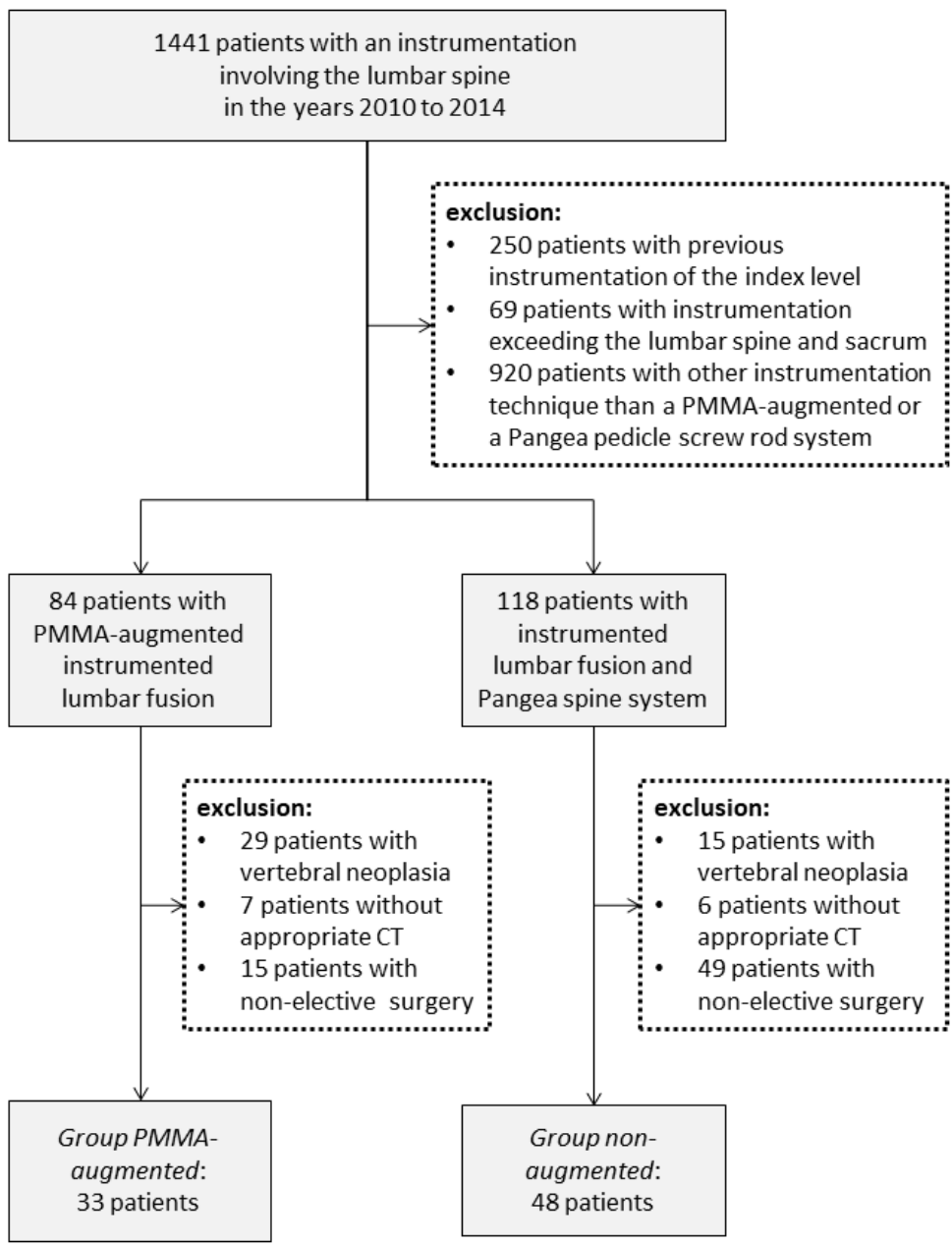

Figure 1: Selection algorithm in the observational study.

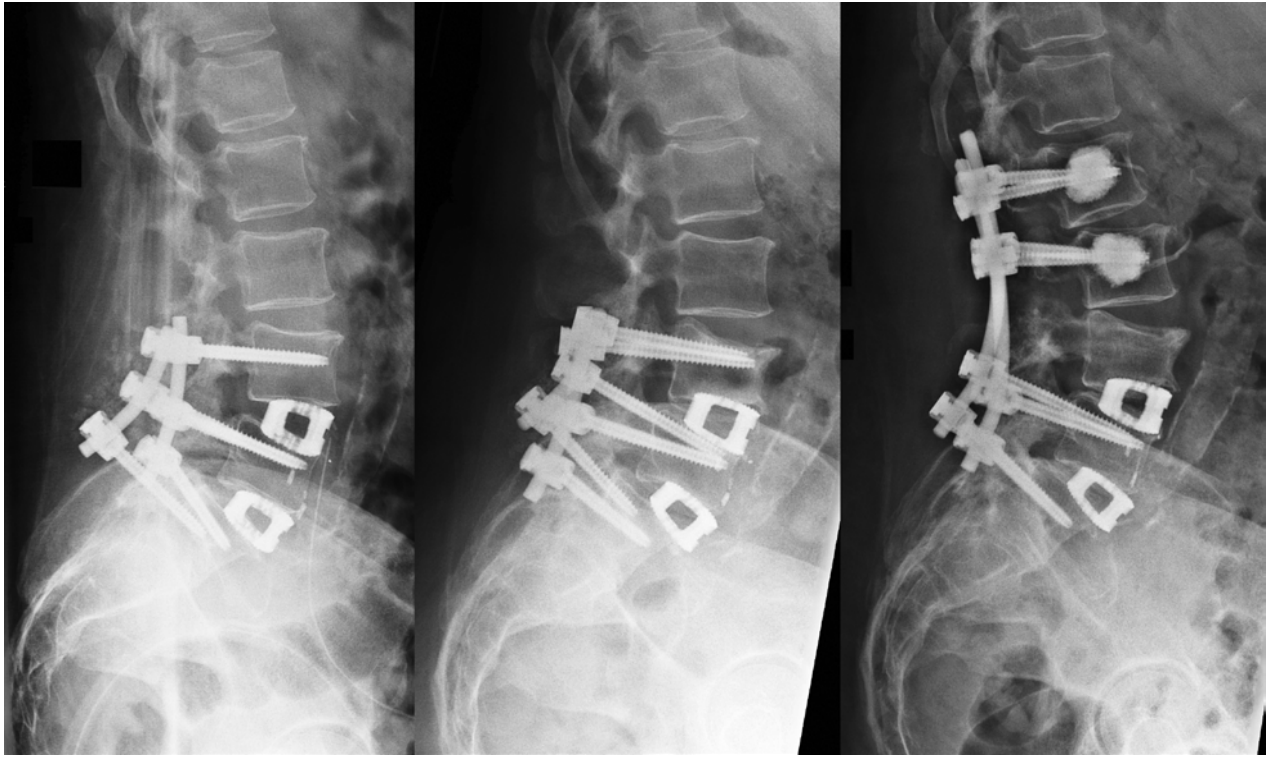

Figure 2: Case of a 76-year old women undergoing instrumented LSF with ALIF for persistent low back pain because of degenerative instability. BMD was not evaluated preoperatively – our retrospective measurement yielded severely osteoporotic BMD of 57 mg/cm³. Left: Initial non-augmented fusion of vertebral levels L4 to S1. Center: Incidence of a compression fracture of the upper instrumented vertebra 22 months later. Right: Reoperation with extended and PMMA-augmented instrumentation of levels L2 and L3; pedicle screws in the fractured vertebra L4 were removed. ALIF, anterior lumbar interbody fusion; PMMA, polymethyl methacrylate.